\documentclass[%
amsmath,amssymb,
aps, physrev,
pre,
twocolumn,
]{revtex4-2}

\usepackage{graphicx}
\usepackage{dcolumn}
\usepackage{bm}

\usepackage[colorlinks=true,linkcolor=blue,citecolor=blue,urlcolor=blue]{hyperref}

\begin{document}

\preprint{APS/123-QED}

\title{Boundary conditions for the Schr{\"o}dinger equation\\ in the numerical simulation of quantum systems}
\thanks{Published in Phys. Rev. E 50, 1616 (1994);
    doi: \href{https://doi.org/10.1103/PhysRevE.50.1616}{10.1103/PhysRevE.50.1616}.
    This preprint incorporates the Erratum, correcting a sign error, published in  Phys. Rev. E 112, 059901(E) (2025); 
    doi: \href{https://doi.org/10.1103/b6qb-59f6}{10.1103/b6qb-59f6}.
}

\author{Marco Patriarca}
\thanks{
Present Address: Academic Visitor, Department of Computer Science, Aalto University.
Email: \tt{marco.patriarca@aalto.fi}; \; \tt{marco.patriarca@gmail.com}; \;
ORCID: \href{https://orcid.org/0000-0001-6743-2914}{0000-0001-6743-2914};\; 
\href{https://scholar.google.com/citations?user=hLUmTDcAAAAJ&hl=en&oi=ao}{Google Scholar profile};\;
\href{https://sites.google.com/view/marco-patriarca}{Home Page}.
}
\affiliation{Istituto Nazionale di Fisica Nucleare and Istituto Nazionale di Fisica della Materia,\\ 
Dipartimento di Fisica, Universit{\`a} di Perugia, Via A. Pascoli 6, 06100 Perugia.}


\begin{abstract}
We study the problem of the boundary conditions in the numerical simulation of closed and open quantum systems, described by a Schrödinger equation. On one hand, we show that a closed quantum system is defined by local boundary conditions. On the other hand, we argue that, because of the uncertainty principle, no local boundary condition can be defined for open quantum systems. For this reason plane waves or wave packet trains cannot be simulated on a finite numerical lattice with the usual procedures. We suggest a method that avoids these difficulties by using only a small numerical lattice and maintains the correspondence with the physical picture, in which the incident and scattered waves may be infinitely extended.
\end{abstract}

\maketitle

\section{\label{sec:intro} Introduction}

Numerical study of the Schrödinger equation, through the solution of a finite-difference equation, is a valuable tool for simulating the dynamics of a quantum system [1]. 
It is particularly useful for the study of transient phenomena when time-dependent or nonlinear effects, as in the Hartree-Fock approximation, have to be taken into account [2].

Moreover, numerical simulations greatly help visualization of quantum mechanics [3] and allow a detailed study of interesting quantum phenomena, which would be very difficult to approach otherwise, such as the``double hump effect'' [4], the loss of coherence in nonquadratic potentials [5], and chaotic behavior in nonlinear tunneling [6]. However, in any numerical simulation one has to specify the boundary conditions, which depend on the particular kind of system under consideration.

In Sec. III we begin with the study of closed quantum systems and show that they can be defined by simple local conditions. By ``local'' we mean that they only involve one point of the lattice.

For open quantum systems, the correct boundary conditions can easily be found in the study of wave-packet scattering [1-4]. In this case the wave function is appreciably different from zero only over a limited region of space, and the situation is equivalent to one of a closed system.

However, it is not always possible to use wave packets. For instance, they are not always good substitutes for plane waves. On one hand, a plane wave of wave vector $k$ can be approximated by a wave packet with average wave vector $k$ if its width $\Delta x$ is much larger than the wavelength $\lambda=2 \pi / k$ of the plane wave, that is, if the condition 
$\Delta x \gg \lambda = 2 \pi / k$ 
holds. 
On the other hand, the dimensions of the numerical grid are limited by the speed and the storage of the computer. If the value of $k$ is too small, the dimensions of the numerical grid, which grow as $1 / k$, are prohibitive. There is unfortunately no way in which, by using some kind of local boundary conditions, a plane wave may be made to enter the system. There is no technical reason for this; it is simply a consequence of the wave nature of the system. The uncertainty principle is directly involved. For example, if we could define a hypothetical ``plane wave injected at a certain point $x=\bar{x}$'', or through a certain interval $\Delta l$ around $\bar{x}$, the product of the position and momentum uncertainties of the incident particles would be exactly zero.

However, as shown in Sec. IV, with a simple modification of the finite-difference equation one can inject a wave of arbitrary shape at a precise point of the lattice. The corresponding physical picture, however, is that in which the wave arrives from an infinite distance. This method is suitable for the simulation of plane waves, wave packets trains, or very large wave packets.

As a check on and illustration of the method, some numerical results are presented in Sec. V. They have been obtained from the numerical solution of the Schrödinger equation (on a two-dimensional $t$-$x$ lattice) with the Crank-Nicolson implicit difference method [7], as explained in detail in the appendix.

\section{\label{sec:problem} STATEMENT OF THE PROBLEM}

We will study a model system, described by a one-dimensional Schrödinger equation [8]
\begin{align}
    i \hbar \frac{\partial \psi(x, t)}{\partial t}
    =
    -\frac{\hbar^{2}}{2 m} \frac{\partial^{2} \psi(x, t)}
        {\partial x^{2}} + V(x, t) \psi(x, t) \, ,
\end{align}
where $V(x, t)$, the external potential, is a general function of the space variable $x$ and of the time variable $t .$ For convenience we introduce an arbitrary unit of energy $\varepsilon_{u}$ and corresponding space and time units $x_{u}$ and $t_{u}$ defined
as $x_{u}=\hbar / \sqrt{2 m \varepsilon_{u}}$ and $t_{u}=\hbar / \varepsilon_{u}$, respectively. Then Eq.
(1) reduces to
\begin{align}
    i \frac{\partial \psi}{\partial t}
    =
    -\frac{\partial^{2} \psi}{\partial x^{2}} + V \psi \, .
\end{align}
If $\varepsilon_{u}=m e^{4} / \hbar^{2}$, then atomic units are obtained. In order to integrate Eq. (2) by numerically solving a finite-difference equation, one has first to define a twodimensional $t$ - $x$ lattice [7].
However, for the sake of clarity, only the $x$ variable will be discretized, while the time variable $t$ will be left continuous. This simplification is possible because none of the following considerations depends on the way the time variable is discretized.

We thus introduce a lattice made of $N+1$ points with coordinates $x_{j}=x_{0}+j \Delta x$, where $j=0, \ldots, N$ and $\Delta x$ is the step of integration. 
Approximating the second derivative with a second difference, from (2) we obtain
\begin{align}
  i \frac{\partial}{\partial t} \psi_{j}
  =
  -\frac{1}{\Delta x^{2}}\left(\psi_{j+1}+\psi_{j-1}-2 \psi_{j}\right)+V_{j} \psi_{j}  \, ,
  \nonumber \\
  j=1, \ldots, N-1,
\end{align}
where $\psi_{j}\left(V_{j}\right)$ represents the wave function (the potential) calculated at time $t$ and at $x=x_{j}$.

However, the system of equations (3) for $j=1, \ldots, N-1$, still cannot be solved until the values $\psi_{0}(t)$ and $\psi_{N}(t)$ have been given. They must be specified through the boundary conditions, in a way that depends on the nature of the physical system to be studied (e.g., a closed or an open one). This will be done in Sec. III for the case of closed quantum systems and in Sec. IV for the case of open quantum systems.

\section{\label{sec:closed} CLOSED QUANTUM SYSTEMS}

A quantum system, described by a wave function $\psi(x, t)$, can be considered closed if a potential $V(x, t)$ is present, such that the conservation law,
\begin{align}
\frac{\partial}{\partial t} \int_{a}^{b} d x|\psi(x, t)|^{2}=0 \, ,
\end{align}
holds at any time $t$. The coordinates $x=a$ and $x=b$ represent two points at a finite distance from each other. Equation (4) holds if the potential at $x=a$ and $x=b$ is large enough to prevent the wave function from escaping. In this case we can assume [9] that
\begin{align}
\psi(a, t) \equiv \psi(b, t) \equiv 0
\end{align}
for any time $t .$ Then the quantum current $J(x, t)$, given by
\begin{align}
J=\frac{\hbar}{2 i m}\left[\psi^{*} \frac{\partial \psi}{\partial x}-\frac{\partial \psi^{*}}{\partial x} \psi\right]
\end{align}
(6)
is zero at $x=a$ and $x=b$, because it is proportional to the local value of the wave function. since no flow is present at the boundaries, from the continuity equation, $\partial|\psi|^{2}/\partial t + \partial J / \partial x=0$, Eq. (4) follows. The boundary conditions for a closed quantum system are naturally suggested by Eq. (5). They turn out to be local and are easily implemented in numerical calculations if we choose the two extreme points of the lattice as
$x_{0}=a$ and $x_{N}=b$,
\begin{align}
\begin{array}{l}
\psi_{0}=0, ~~~ \psi_{N}=0
\end{array}
\end{align}
In the case of a harmonic oscillator, for example, $x_{0}$ and $x_{N}$ can be chosen as two points where the potential is very large (much larger than the average energy of the wave function).

An interesting example is given by an external potential which is zero everywhere, that is, $V(x, t) \equiv 0 $. 
In this case, what system do conditions (7) represent? 
Since they are equivalent to the assumption that a very large potential is present at $x=x_{0}$ and $x_{N}$, then $V$ increases sharply at $x=x_{0}$ and $x_{N} .$ This is equivalent to a particle moving between perfectly reflecting walls. 
For this problem the exact solution is known, both with one and two reflecting walls, if the initial shape of the wave function is Gaussian. 
This allows a direct comparison between numerical and analytical results in order to verify the validity of conditions (7).

It is to be noted that conditions (7) imply the conservation of the total probability $\int d x|\psi(x, t)|^{2} .$ Its finite-difference counterpart is $\Delta x \mathbf{\Sigma}_{j}\left|\psi_{j}\right|^{2} .$ It we take its time derivative, by making use of (3) and (7) and assuming that the potential $V$ is real, we obtain
\begin{align}
    \frac{\partial}{\partial t} \Delta x \sum_{j=1}^{N-1}\left|\psi_{j}\right|^{2} 
    & = \Delta x \sum_{j=1}^{N-1} \left( 
              \frac{\partial\psi_{j}^{*}}{\partial t}\psi_{j}
            + \psi_{j}^{*}\frac{\partial\psi_{j}}{\partial t}\right) 
        \nonumber \\
    & = \frac{2}{\Delta x} \operatorname{Im} (\psi_{0}^{*} \psi_{1}+\psi_{N}^{*} \psi_{N-1} ) 
        \equiv 0
\end{align}
%

\begin{center}
\begin{figure}[t!]
\centering
\includegraphics[scale=1.2]{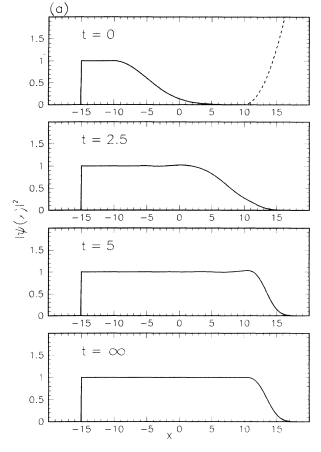} 
\includegraphics[scale=1.35]{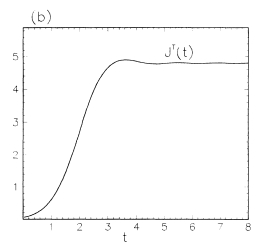} 
    \caption{Propagationof the plane wave $\Phi_{0}(x, t) = A \exp \left(i k x-i k^{2} t\right)$,
    with $A=1$ and $k=2.4$, injected at $x_{s}=-15$.
    (a) $P(x, t)=|\psi(x, t)|^{2}$ vs $x$ at $t=0,2.5,5, \infty$
    (by `` $\infty$'' we mean a large time $t \simeq 20$).
    The  initial configuration is
    $\psi(x, 0)=0$ for $x \leq x_{s}$ and
    $\psi(x, 0)=\Phi_{0}(x, 0) g(x)$ for $x>x_{s}$,
    where $g(x)$ is given by (9).
    The dashed line represents the absolute value (in arbitrary units) of the imaginary potential $i V_{i}(x)=-i c\left(x-x_{i}\right)^{2} \Theta\left(x-x_{i}\right)$, where $\Theta$ is the Heaviside function, $x_{i}=20$, and $c=0.1$, plotted only at $t=0$ for clarity. (b) Quantum current calculated numerically from Eq. (6) at $x=10$.}
\label{Fig_algo}
\end{figure}
\end{center}

\section{\label{sec:open}OPEN QUANTUM SYSTEMS}

We commence with the simpler problem of producing a wave $\Phi_{0}(x, t)$ at a certain point $x=x_{s}$ ($s$ for source) traveling toward the right, e.g., a plane wave $\Phi_{0}(x, t)=A \exp \left(i k x-i k^{2} t\right)$ with $k>0 .$ 
For the moment we neglect any possible interactions [that is, we assume $V(x, t) \equiv 0]$ which would produce reflected waves. 
To solve this problem the following steps are sufficient:
(i) Consider the points $x_{j}$ with $j \geq s$ and neglect the rest of the lattice. (ii) Put $\psi_{s}(t) \equiv \Phi_{0}\left(x_{s}, t\right)$
(iii) Add a negative imaginary potential near the right border to absorb the wave, so that at $x=x_{N}$ one has $\psi_{N}(t)=0$
(iv) Solve the system of Eq. $(3)$, for $j=s+1, \ldots, N-1$ with the boundary conditions given by (ii) and (iii) (note that they are local). 

As examples of initial conditions, we choose 
$\psi\left(x_{j}, 0\right) = \Phi_{0}\left(x_{j}, 0\right) g\left(x_{j}\right)(j>s)$, 
where $\Phi_{0}$ is the plane wave. 
The functions $g(x)$ describes the shape of the initial wave function for $x>x_{s}$.

If we take $g(x) \equiv 1$, a plane wave is observed to be continuously generated at $x=x_{s}$ and absorbed near the right border. 
If $g(x)$ varies smoothly from $g=1$ to $g=0$ as $x$ increases, a wave front is observed moving toward the right. Figure $1(a)$ shows plots of the probability density $|\psi(x, t)|^{2}$ versus $x$ different instants of time. They have been obtained from a numerical simulation in which a plane wave $\Phi_{0}(x, t)=A \exp \left(i k x-i k^{2} t\right)$, with $A=1$ and $k=2.4, \quad$ is $\quad$ produced $\quad$ at $\quad x_{s}=-15 . \quad$ The initial configuration at $t=0$ is assumed to be of the form $\psi(x, 0)=\Phi_{0}(x, 0) g(x), \quad$ where $\quad g(x)$ corresponds to a smooth wave front and is given by
\begin{align}
    g(x) 
    =  
    \left\{
        \begin{array}{l}
            1 \text{~~~~for~~} x_{s} < x <x_{g} \, , \\
            \exp\left[-\left(x-x_{g}\right)^{2}/l_{g}^{2}\right] \text{~~for~~} x > x_{g} \, ,
        \end{array}
    \right.
\end{align}
where $x_{g}=-10, l_{g}=3 .$ 
The wave front is observed moving toward the right with some small oscillations and with a velocity given by $\left.v=2 k \text { (in the units } \varepsilon_{u}, x_{u}, t_{u}\right)$ Figure $1(b)$ shows the quantum current $J(t)$, given by (6) calculated numerically at a point $x>x_{s}$ versus time $t$ For large values of $t, J(t) \rightarrow J_{0}=2 k|A|^{2}$, the incident current.

The oscillations can be unstable if the absorbing potential is too far from the point of injection or if $g(x)$ is not smooth enough. An example of unstable initial conditions is given by $\psi(x, 0) \equiv 0$, which can be obtained with $g(x) \equiv 0 .$ Even if it looks smooth $(\psi \equiv 0$ everywhere), it is physically equivalent to the highly discontinuous initial conditions $\psi(x, 0)=\Phi_{0}(x, 0) \Theta\left(x-x_{s}\right)$, where $\Theta$ is the Heaviside function. Actually the plane wave is also present in the region $x \leq x_{s}$, even if in this part of the lattice it is hidden.

After a transient, a steady state is reached, in which the plane wave is continuously produced at $x=x_{s}$ and absorbed near the right hand border by the imaginary potential $V_{i}(x)=-i c\left(x-x_{i}\right)^{2} \Theta\left(x-x_{i}\right)$, where $c=0.1$ and $x_{i}=20 .$ A smooth imaginary potential produces a gradual absorption of the wave and avoids oscillations, which could propagate back through the whole lattice. Moreover, in this way there is a large saving of calculations because one does not have to follow the evolution of the wave.

\begin{center}
\begin{figure}[t!]
\centering
\includegraphics[scale=1.2]{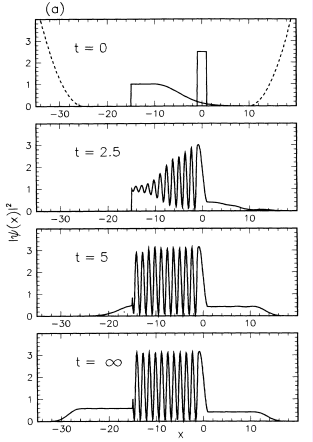} 
\includegraphics[scale=1.4]{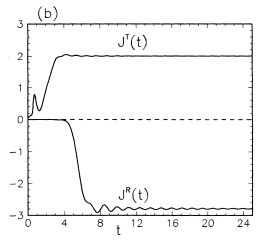} 
    \caption{Same wave of Fig. $1$, scattering off the square potential barrier
$V(x)=V_{0} \Theta(x-a) \Theta(x-b)$,
where $V_{0}=5, a=-1$ $\boldsymbol{b}=+1$.
(a) $|\psi(\boldsymbol{x}, t)|^{2}$ at $t=0,2.5,5, \infty .$
For the sake of clarity, the real potential $V(x)$ (continuous line) and the absolute value of the imaginary potential $V_{i}(x)$ (dashed line) are plotted only at $t=0$.
The oscillations between the point of injection and the barrier are due to the interference between incident and reflected waves.
(b) Reflected (negative) current $J^{R}(t)$ and transmitted (positive) current $J^{T}(t)$, evaluated at $x=-20$ and $10$, respectively.}
\label{Fig_algo}
\end{figure}
\end{center}

We now consider the more general situation in which scattering of the incident wave $\Phi_{0}(x, t)$ takes place. We assume that a general potential $V(x, t)$ is present for $x>x_{s}$ and an initial configuration $\psi(x, 0)$ has been assigned. We expect a reflected wave $\Psi^{R}(x, t)$ and a transmitted wave $\Psi^{T}(x, t)$ to appear, whose shapes are not known in advance. As far as the transmitted wave is concerned, no restriction on $V$ is required for its definition. Its study can be comfortably carried out in the region between the potential, where it is generated, and the right hand border, where it is absorbed by the imaginary potential.

On the other hand, the definition of the reflected wave
$\Psi^{R}$ rests on the assumption that $V(x, t)$ becomes constant for $x \rightarrow-\infty$ (if the incident wave comes from the left), so that the total wave function there can be written as a superposition $\Psi^{\text {Tot }}=\Psi^{R}+\Phi_{0}$. This is why it has been assumed that $V \rightarrow 0$ in the left hand region of the lattice (namely that $V \equiv 0$ for $x \leq x_{s}$).

The reflected wave $\Psi^{R}$ will propagate from the potential region toward the left, superposed on the incident wave $\Phi_{0} .$ After it has reached the point $x=x_{s}$, the lattice points $x_{j}$, with $j<s$, have to be taken into account. However, this must be done in such a way that the apparent discontinuity of the wave function at $x=x_{s}$, due to the production of $\Phi_{0}$, does not propagate to the left as if it were a physical discontinuity.

If one considers Eq. (3) for fixed $j$, one notices that it only contains values of $\psi$ as three different points of the lattice [10] (at the points $j+1, j, j-1$). 
Then if in Eq. (3), written at the point $j=s$, the value of $\Phi_{0}$ is subtracted from $\psi_{j+1}$, everything goes as if ``the point $x=x_{s}$ only sees the reflected wave $\Psi^{R}$'', and by applying (3) at all points $x_{j}$, with $j<s$, the only reflected wave evolves.

In the same manner, if in Eq. (3), written at the point $j=s+1$, the value of $\Phi_{0}$ is added to $\psi_{j-1}$, everything goes as if ``the point $x=x_{s+1}$ sees the total wave function $\Psi^{\mathrm{Tot}}=\Psi^{R}+\Phi_{0}$'' . 
Thus it is $\Psi^{\mathrm{Tot}}$ that evolves in the rest of the lattice by just applying Eq. (3) at all the points $j>s$. In terms of formulas, the production of the incident wave at $x=x_{s}$ and the evolution of the incident and scattered waves are carried out through the following steps.

(i) Apply (3) to all points $x_{j}$ with $1 \leq j \leq s-1$.
Here $\psi$ only represents $\Psi^{R}$.

(ii) At $j=s$ subtract $\Phi_{0}$ from $\psi_{j+1}$ 
(because $\psi_{s} = \Psi_{s}^{R}$ but 
$\psi_{s+1} = \Psi_{s+1}^{\mathrm{Tot}}$). 
Then Eq.~(3) becomes
\begin{align}
\nonumber
i \frac{\partial \psi_{s}(t)}{\partial t} 
    = &- \frac{[\psi_{s+1}(t) - \Phi_{0}(x_{s+1}, t) ]  + \psi_{s-1}(t) - 2 \psi_{s}(t)}{\Delta x^{2}} \\
      &+ V_{s}(t) \psi_{s}(t) \, .
\end{align}

(iii) At $j=s+1$ add $\Phi_{0}$ to $\psi_{j-1}$ 
(because $\psi_{s+1} = \Psi_{s+1}^{\mathrm{Tot}}$ but 
$\psi_{s} = \Psi_{s}^{R}$),
\begin{align}
\nonumber
i \frac{\partial \psi_{s+1}(t)}{\partial t}
    = &- \frac{ \psi_{s+2} + [\psi_{s}(t) + \Phi_{0}(x_{s}, t) ] - 2 \psi_{s+1}(t) }{\Delta x^{2}}\\ 
      &+ V_{s+1}(t) \psi_{s+1}(t) \, .
\end{align}

(iv) Apply (3) to the rest of the points $x_{j}$, with $j \ge s+2$, 
here $\psi = \Psi^{\text{Tot}}$.

(v) Put $\psi_{0}(t) \equiv 0$ and $\psi_{N} \equiv 0$ at the left and right hand boundaries, where two suitable imaginary potentials absorb the reflected and transmitted waves, respectively. 

The finite-difference system, defined by (i), (iii), (iii), and (iv), with the boundary conditions given by (v), can now be solved by standard methods. 
This procedure is valid not only when $\Phi_{0}(x, t)$ is a plane wave, but for any other incident wave (e.g., superposition of plane waves, trains of wave packets, or just a single wave packet whose width is very large). 
It is also applicable to nonlinear potential, for instance to self-consistent Schrödinger equations, where $V(x, t,[\psi])$ is a functional of the wave function $\psi$. 
The only condition is that $V=0$ (or constant) for $x<x_{s}$.
The calculation of the reflected wave, through the subtraction of the incident wave from the total wave function, was first carried out by Mains and Haddad [11] in a numerical study of resonant-tunneling diodes. 
With the additional condition that $V(x, t) \simeq const$ for $x \leq x_{s}$, it is possible to study reflected and transmitted waves for a wide category of quantum mechanical systems, using the method illustrated in this section.

\begin{center}
\begin{figure}[t!]
\centering
\includegraphics[scale=1.5]{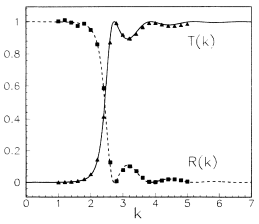} 
    \caption{Reflection (dashed line) and transmission (continuous line) coefficients $T$ and $R$ of the square potential barrier of Fig. 2(a), calculated from Eqs. (12) and (13), are compared with the values obtained by numerical simulations:
    numerical values of $R$ ($\blacksquare$) and numerical values of $T$ (``$\blacktriangle$'').
    The integration steps used in the numerical simulation are $\Delta t=0.01$ and $\Delta x=0.05$.}
\label{Fig_algo}
\end{figure}
\end{center}

\section{\label{sec:numerics}NUMERICAL SIMULATIONS}

Finally, we present some numerical examples. The numerical results have been obtained using the Crank-Nicolson implicit method [7], as described in detail in the Appendix.

We begin with the study of a plane wave 
$\Phi_{0}(x, t) = A \exp \left(i k x-i k^{2} t\right)$ 
with $A=1$, $k=2.4$, produced at $x_{s}=-15$ and scattered by a static square barrier 
$V(x) = V_{0} \Theta(x-a) \Theta(b-x)$, with $V_{0}=5, a=-1$, $b=+1$. 
Figure $2(a)$ shows $|\psi(x, t)|^{2}$ versus $x$ at different times. 
One can see how the incident plane wave scatters off the barrier, producing a transmitted and a reflected wave. 
The reflected wave interferes with the incident one between the point of injection and the potential region, giving rise to fast oscillations of $|\psi|^{2}$. 
They are localized in the region between $x=x_{s}$ and the barrier, where both the incident and the reflected wave are present. 
The wavelength of the oscillations is $\lambda =\frac{1}{2}(2 \pi / k)=\pi / k$ 
(the factor $\frac{1}{2}$ is due to the fact that the second power of $|\psi|$ is plotted). 
For $x<x_{s}$ no interference is observed, because $\Phi_{0}$ has been subtracted from 
$\Psi^{\text {Tot }}$ at $x=x_{s}$ and only $\Psi^{R}$ has been left. 
The reflected wave is then visualized in the region $x<x_{s}$ and its detailed study is carried out just as for the transmitted wave in the region behind the barrier. 
For this reason the absence of interference turns out to be very useful, even if nonphysical --in a one-dimensional system there is always interference between incident and reflected waves.

\begin{center}
\begin{figure}[t!]
\centering
\includegraphics[scale=1.5]{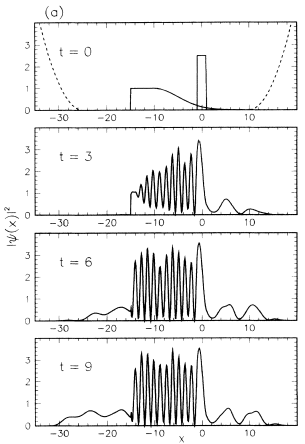} 
\includegraphics[scale=1.5]{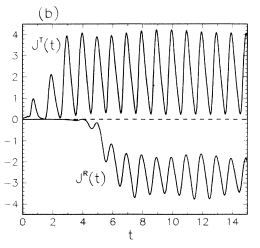} 
    \caption{Same plane wave of Figs. 1 and $2$, scattering off the time-dependent potential
    $V(x, t) = V_{0} [1 + \alpha \cos (\omega t)] \Theta(x-a) \Theta(b-x)$,
    where $V_{0} = 5$, $\alpha = 1/2$, $v = \omega / 2 \pi = 1$, $a = -1$, $b = +1$.
    (a) $|\psi(x, t)|^{2}$ at $t=0,3,6,9$.
    Note the oscillations in space.
    (b) Reflected and transmitted currents, calculated at $x=-20$ and $x=10$, respectively.
    Notice the oscillations in time.}
\label{Fig_algo}
\end{figure}
\end{center}

Figure 2(b) shows the time evolution of the transmitted and reflected currents $J^{T}(t)$ and $J^{R}(t)$. 
It can easily be shown that for $t \rightarrow \infty$ they approach just the values $2 k|A|^{2} T(k)$ and $2 k|A|^{2} R(k)$ corresponding to the stationary Schrödinger problem [12], where $R(k)$ and $T(k)$ are the reflection and the transmission coefficients. This means that the system relaxes toward its stationary state. Figure 3 shows a comparison between the values of the transmission and reflection coefficients obtained by numerical simulation and those calculated from the theoretical expressions for the case of a square potential barrier [12], given by
\begin{align}
T(k) = 1-R(k) = 
    \frac{1}{1+\frac{1}{4}
            \left[ \frac{k^{\prime}}{k}-\frac{k}{k^{\prime}}\right]^{2}
            \left[\sin \left(k^{\prime} L\right)\right]^{2}}
\end{align}
for $k^{2}>V_{0}$ and by
\begin{align}
T(k)=1-R(k)=\frac{1}{1+\frac{1}{4}\left(\frac{k^{\prime}}{k}+\frac{k}{k^{\prime}}\right)^{2}\left[\sinh \left(k^{\prime} L\right)\right]^{2}}
\end{align}
for $k^{2}<V_{0} .$ Here $k^{\prime}=\sqrt{\left|V_{0}-k^{2}\right|}$ and $L=b-a$ is the barrier length.
As a last example, Figs. $4(a)$ and $4(b)$ show the results of a numerical simulation of a time-dependent problem:
the scattering of a plane wave by a square barrier whose height oscillates in time. Even for systems of this kind, the present method allows us to study directly the time evolution of the wave function with no approximation other than that due to the numerical algorithm used.

\section{\label{sec:conclusion } Conclusion}

It has been shown that numerical simulations of closed quantum systems can be carried out by using local boundary conditions for the Schrödinger equation. On the other hand, however, the concept of boundary condition is meaningless for open quantum systems because of the uncertainty principle. In this case the problem could be avoided by using a lattice which is large enough to contain the shape of the whole function. But if one has to study very large wave packets or plane waves, this is not a realistic solution.

We have shown, however, that it is possible to provide a numerical description of an open quantum system, where incident as well as reflected and transmitted waves are present. The method to be used produces the incident wave at a certain point of the lattice. The transmitted wave is studied behind the interaction region, while the reflected wave is isolated by extracting the incident wave from the total function at the point of injection. All this is obtained with some simple modifications of the finite-difference equation used. The study of the evolution of the transmitted and reflected waves very far from the scattering region is avoided by imaginary potentials which absorb the scattered waves. The corresponding physical picture is, however, meaningful because it describes waves which may be extended infinitely.

This approach to the numerical study of scattering phenomena could be useful for studying problems where time dependence is an essential ingredient (e.g., transient phenomena and scattering by time-dependent potentials) without either resorting to perturbation theory, partial wave analysis, or using wave packets.


\section*{ACKNOWLEDGMENTS}

I wish to thank Carlo Presilla, University of Rome ``La Sapienza,'' Italy, for many invaluable suggestions and discussions, and Roberto Onofrio, University of Padova, Italy and Derek Boothman, University of Perugia, Italy, for a careful reading of the paper. The numerical calculations have been carried out with the computers of the INFN, Istituto Nazionale di Fisica Nucleare, Sezione di Perugia, Italy.

\section*{References}
\noindent
[1] For recent applications, 
see B. Ritchie, Phys. Rev. A 45 4207(1992), and references therein.
\vspace{0.5ex}  

\noindent
[2] G. Jona Lasinio, C. Presilla, and F. Capasso, 
Phys. Rev. B 43, 5201 (1991).
\vspace{0.5ex}  

\noindent
[3] I. Galbraith, Y. S. Ching, and E. Abraham, 
Am. J. Phys. 52 , 61 (1984), and references therein.
\vspace{0.5ex}  

\noindent
[4] M. H. Bramhall and B. M. Casper, 
Am. J. Phys. 38, 1136 (1970).
\vspace{0.5ex}  

\noindent
[5] M. M. Nieto, in 
\textit{Coherent States}, 
edited by J. R. Klauder (World Scientific, singapore, 1985).
\vspace{0.5ex}  

\noindent
[6] G. Jona Lasinio, C. Presilla, and F. Capasso, 
Phys. Rev. Lett. 68,2269(1992).
\vspace{0.5ex}  

\noindent
[7] W. H. Press, B. P. Flannery, S. A. Teukolsky, and w. T. Wetterling, 
\textit{Numerical Recipes} (Cambridge University Press, Cambridge, 1986).
\vspace{0.5ex}  

\noindent
[8] We do not consider open systems coupled to thermal baths, for which different approaches are required. 
Compare W. R. Frensley, Rev. Mod. Phys. 62, 745 (1990).
\vspace{0.5ex}  

\noindent
[9] L. I. Schiff, 
\textit{Quantum Mechanics}, 3rd ed. (McGraw-Hill, New York, 1968), Chap. 2.8.
\vspace{0.5ex}  

\noindent
[10] It is not difficult to generalize the following considerations to finite-difference equations that depend on more than three different points.
\vspace{0.5ex}  

\noindent
[11] R. K. Mains and G. I. Haddad, 
J. Appl. Phys. 64, 3564 (1988).
\vspace{0.5ex}  

\noindent
[12] L. I. Schiff,  3rd ed. (McGraw-Hill, New York, 1968), Chap. 2.8,
\textit{Quantum Mechanics}, Chap. 5.17.
\vspace{0.5ex}  

\noindent
[13] M. Patriarca, 
\textit{Boundary conditions for the Schr{\"o}dinger equation in the numerical simulation of quantum systems},
Phys. Rev. E 50(2), 1616 (1994).
DOI: \href{https://doi.org/10.1103/PhysRevE.50.1616}{10.1103/PhysRevE.50.1616}
\vspace{0.5ex}  
 
\noindent
[14] M. Patriarca, 
\textit{Erratum: Boundary conditions for the Schr{\"o}dinger equation in the numerical simulation of quantum systems [Phys. Rev. E {\bf 50}, 1616 (1994)]},
Phys. Rev. E 112, 059901(E) (2025).
DOI: \href{https://doi.org/10.1103/b6qb-59f6}{10.1103/b6qb-59f6}
\vspace{0.5ex}  

\noindent
[15] While in the original paper [13] the expressions for $\beta_{j}, \overline{\beta}_{j}$ contain a sign error in the potential terms (eventually corrected in the erratum [14]), Eqs. (A5)-(A7) here have the right signs.



\appendix

\section{Numerical scheme}

We have used the Crank-Nicolson implicit finite-difference method [7] to discretize Eq. (3) on a two-dimensional lattice $\left(t_{k}, x_{j}\right)$, with $t_{k}=k \Delta t(k=0, \ldots, M)$ and $x_{j}=x_{0}+j \Delta x(j=0, \ldots, N)$. 
$\Delta t$ and $\Delta x$ are the integration steps for the $x$ and the $t$ variables, respectively. 
We refer to the values $\psi(t_{k}, x_{j}) [V(x_{j}, t_{k})]$  as $\psi_{j}^{k}(V_{j}^{k})$.
Then the finite-difference equation corresponding to the Crank-Nicolson implicit method can be obtained through the following simple time discretization procedure: 

-- the derivative $\partial \psi_{j} / \partial t$ is replaced with the finite difference 
$\left(\psi_{j}^{k+1}-\psi_{j}^{k}\right) / \Delta t$ 

-- the values of $\psi_{j}\left(V_{j}\right)$ are replaced with their arithmetic averages between times $t_{k}$ and $t_{k+1}$,
$\frac{1}{2} \left( \psi_{j}^{k} + \psi_{j}^{k} \right) \left[ \frac{1}{2} \left( V_{j}^{k+1} + V_{j}^{k} \right) \right]$,
\begin{align}
  &\psi_{j}^{k+1}-\psi_{j}^{k} 
  \nonumber \\
  &= \frac{i \Delta t}{2 \Delta x^{2}} \!
            \left[\!\left(\psi_{j+1}^{k+1} \!+\psi_{j-1}^{k+1}-2 \psi_{j}^{k+1}\right)
            \!+\!
            \left(\psi_{j+1}^{k}+\psi_{j-1}^{k}-2 \psi_{j}^{k}\right)\!\right] 
  \nonumber \\
  &~~~~+i \Delta t ~\frac{V_{j}^{k+1}+V_{j}^{k}}{2} ~ \frac{\psi_{j}^{k+1}+\psi_{j}^{k}}{2} \, .
\end{align}
The integration steps used in the present paper are $\Delta t \simeq 0.01$ and $\Delta x \simeq 0.05$.
We point out that if we want the quantity
$\int dx \, |\psi(x, t)|^{2}$ to be conserved, then the time average of the term $V_{j} \psi_{j}$ in (3) must be done in such away that the coefficients of $\psi_{j}^{k+1}$ and of $\psi_{j}^{k}$ in the last term of (A1) are equal. 
In this case, by making use of (A1) and letting $\psi_{0}^{k} \equiv \psi_{N}^{k} \equiv 0$, it is possible to show that
\begin{align}
    \sum_{j=1}^{N-1} \left|\psi_{j}^{k+1}\right|^{2}
    =\sum_{j=1}^{N-1}\left|\psi_{j}^{k}\right|^{2} \, .
\end{align}
For other kinds of time averaging, such as $V_{j} \psi_{j} \rightarrow \frac{1}{2}\left(V_{j}^{k+1} \psi_{j}^{k+1}+V_{j}^{k} \psi_{j}^{k}\right)$, the time evolution operator associated with the finite-difference equation is not unitary and the quantity $\int d x|\psi(x, t)|^{2}$ is not conserved.
If $V$ does not depend on $\psi$, Eq. $(A 1)$ can be reordered in the form of a tridiagonal system, 
\begin{equation}
    \alpha \, \psi_{j+1}^{k+1} + \beta_{j} \, \psi_{j}^{k+1} + \alpha \, \psi_{j-1}^{k+1} = r_{j} \, , 
    \quad j=1, \ldots, N-1.
\end{equation}
The coefficients $\alpha, \beta_{j}$, and $r_{j}$ do not depend on the values of $\psi$ at time $k+1$ and are given by [15]
\begin{align}
    \alpha &= -i \Delta t / 2 \Delta x^{2} \, , \label{A4}\\
    \beta_{j} &= 1 + i \Delta t / \Delta x^{2} + i \Delta t\left(V_{j}^{k+1}+V_{j}^{k}\right) / 2 \, , \label{A5}\\
    r_{j} &= \alpha^{*} \psi_{j+1}^{k} + \bar{\beta}_{j} \psi_{j}^{k} + \alpha^{*} \psi_{j-1}^{k} \, . \label{A6}
\end{align}
where
\begin{align}
\overline{\beta}_{j} = 1 - i \Delta t / \Delta x^{2} - i \Delta t \, (V_{j}^{k+1}+V_{j}^{k}) / 2 \, .
\end{align}
Note that if $V$ is complex, the $\bar{\beta}_{j}$ are not the complex conjugate of the $\beta_{j}$. 

The modifications of the finite-difference equations represented by (10) and (11)
can be taken into account here simply by modifying the terms $r_{s}$ and $r_{s+1}$,
\begin{align}
    &r_{s} \rightarrow r_{s}^{\prime}=r_{s}+\alpha\left(\Phi_{s+1}^{k+1}+\Phi_{s+1}^{k}\right) \, ,
    \\
    &r_{s+1} \rightarrow r_{s+1}^{\prime}=r_{s+1}+\alpha^{*}\left(\Phi_{s}^{k+1}+\Phi_{s}^{k}\right) \, .
\end{align}
System (A3) can be solved using the Gauss elimination method [7] to obtain the $\psi_{j}^{k+1}$'s as functions of the $\psi_{j}^{k}$'s, once the values of $\psi_{0}^{k+1}$ and $\psi_{N}^{k+1}$ are known.

\end{document}